
\input harvmac
\Title {DAMTP93/R32}
{{\vbox {\centerline{Yang-Mills Cosmologies}\break
\centerline{ and
Collapsing Gravitational Sphalerons  }}}}

\bigskip
\centerline{G.W. Gibbons}
\centerline{Alan R. Steif}
 \centerline{\it Department of Applied Mathematics and Theoretical
Physics}
\centerline {\it Cambridge University}
\centerline{\it Silver St. }
 \centerline{\it Cambridge,   CB3 9EW }
\centerline {\it United Kingdom }
\centerline{\it gwg1@amtp.cam.ac.uk}
\centerline{\it ars1001@amtp.cam.ac.uk}
 \vskip .2in

\noindent
ABSTRACT: Cosmological solutions with a homogeneous
Yang-Mills field which  oscillates and passes
between topologically distinct vacua are
discussed. These solutions are used to model the collapsing
Bartnik-McKinnon gravitational sphaleron
and the associated anomalous production of
fermions. The Dirac equation   is analyzed
in these backgrounds. It is shown explicitly
that   a   fermion energy level crosses
from the negative to positive energy spectrum as the
gauge field evolves between
the topologically distinct vacua. The cosmological
solutions are also generalized to include an axion field.

\Date {}
\def\a{\alpha}
\def\b{\beta}
\def\ra{\rightarrow}
\def\o{\omega}
\def\m{\mu}
\def\pa{\partial}
\def\p{\partial}
\def\th{\theta}
\def\t{\tau}
\def\s{\sigma}

\def\l{\lambda}
\def\n{\nu}
\def\g{\gamma}

\def\d{\bf d}

\def\sne{{\rm sin}\, \eta}

\def\ep{\epsilon}

\def\D3{\hbox{$D_3\kern -10pt / \kern 10pt$}}
\def\Aslash{\hbox{$ A^a\kern -10pt / \kern +10pt$}}
\def\paslash{\hbox{$ \partial \kern -6pt / \kern +6pt$}}

\def\P{\Psi}
\def\O{\Omega}
\def\F{F_{\m\n}}
\def\tr {{\rm Tr}}

\def\e{\eta}
\def\be{{\bf e}}
\def\w{\wedge}
\def\r{\rho}
\def\de{\delta}

\def\plb{\it Phys. Lett. B}
\def\cqg{\it Class. Quantum Grav.}
\def\npb{\it Nucl. Phys. B}
\newsec{Introduction}

 A few years ago the Einstein-Yang-Mills field equations
were  shown to admit spherically symmetric particlelike solutions in which the
Yang-Mills configuration
is bound by gravity in unstable equilibrium \ref\bm{
R. Bartnik and J. McKinnon, {\it Phys. Rev. Lett.} {\bf 61} (1988) 141.
}.  These solutions correspond to gravitational analogs of the standard model
sphaleron solution  \ref\gv{
D. Gal'tsov and M. Volkov, {\it Phys. Lett. } {\bf B273} (1991) 255; D.
Sudarsky and R. Wald, {\it Phys. Rev. D} {\bf 46} (1992) 1453.}.
 In other words, they are  saddlepoint solutions that lie on an
energy barrier in field configuration space separating vacua with different
topological
charge. Just as  the ordinary sphaleron  mediates transitions between
these vacua and leads to   anomalous fermion production
so  should its gravitational analog.
 Consider a gauge field which vanishes as $t\ra - \infty$ and is
given by a pure gauge with unit Chern-Simons number as $t\ra \infty$.
According to the anomaly equation,  the change in fermion number is given
by  the change in Chern-Simons number:
$\Delta n = \int F \wedge F d^4x$. Ordinarily, even in a time-dependent
background
particles and anti-particles are created in pairs. However, as discussed
in {\ref\anomaly{ R. Wald, {\it Ann. Phys.} {\bf 118} (1979) 490; G. Gibbons,
{\plb} {\bf 84} (1979) 431; G. Gibbons, {\it Ann. Phys.}
{\bf 125} (1980) 98; N. Christ, {\it Phys. Rev. D} {\bf21} (1980) 1591. }},
this is not necessarily the case for fermions.
  For slowly varying fields,
the
anomalous production of fermions has a simple description in terms of
the spectral flow of the Dirac operator. Consider the spectrum
 in the background of the sequence of instantaneous
static field configurations with the initial and final gauge field
configurations being pure gauge but with their Chern-Simons numbers
differing by $n$. The initial and final spectra should coincide with the
spectrum in the absence of a gauge field. However, the anomalous
production is realized by
  $n$ energy levels   crossing
    from negative to positive values of the energy. Thus,
assuming  one has begun  in the vacuum with the negative energy Dirac sea
filled, afterwards,
there would be $n$ positive energy levels occupied corresponding to the
creation
of $n$ particles. Since the sphaleron configuration is the midpoint in the
transition, one would expect that  there would be a zero energy
normalizable fermion  mode in the sphaleron background. Indeed, this was shown
to be the case
for the Bartnik-McKinnon solution {\ref\us{G. Gibbons and A. Steif, {\it Phys.
Lett. B.} {\bf 314}   (1993) 13.}}.

For the sphaleron of the standard model, one may view the time dependent
process
as being initiated by a high-energy  collision of particles
and ending with the  fields going off to infinity.   For the gravitational
sphaleron,
in addition to this scenario, there is  the possibility of the configuration
undergoing gravitational collapse
and forming a horizon and singularity.
 Thus, the anomalously produced fermions could either enter
the horizon or go off to infinity. [One should note that the
third possiblity of the fermions
  remaining bound  outside the black hole is excluded by the no-hair theorem.]
To calculate
the fraction of fermions that go off to infinity, one should  study the
  evolution of the fermion zero-mode.  Thus, one should solve
  the time-dependent
Dirac equation in the background of
the collapsing   Bartnik-McKinnon solution.
This is of course a difficult   problem for which
little progress can be made analytically.

 In this paper, we study
cosmological
solutions to the Einstein-Yang-Mills equations with closed spatial slices
 with the  motivation that they might serve
as an approximation  for the interior of the collapsing sphaleron.
 Since   $SU(2)$ and
the spatial sections are both three-spheres, there is a natural ansatz for
a homogeneous $SU(2)$ gauge field involving a single   function of time, $f$.
The energy momentum tensor of the gauge field corresponds to pure radiation,
and the spacetime to the Tolman universe. In contrast with
abelian electromagnetism where the only way to obtain a pure radiation
stress tensor is by
some averaging procedure, with  a non-abelian gauge field, one can obtain a
pure radiation stress tensor directly from a classical solution. As we will
discuss, these cosmological
solutions share features of  the sphaleron.  The
time evolution of $f$ corresponds to motion in a double well with minima at the
topologically distinct vacua. There is an  unstable  gauge field
configuration at the top
of the energy barrier  which corresponds  to the
 reduction of the four-dimensional meron
solution
\ref\aff{V. de Alfaro, S. Fubini, and G. Furlan,
{\it Nuovo Cimento}  {\bf 50} (1979) 4523.} to three-dimensions. There is  a
fermion zero mode in the background of this  configuration with the same
internal structure  as  the zero mode  in the
Bartnik-McKinnon solution. Moreover, we show explicitly that as the field
evolves between the topologically distinct vacua there is a single
fermion level crossing corresponding to the anomalous production
of one particle.

In Section 2, we   review   Robertson-Walker spacetimes
with a homogeneous Yang-Mills field. In Section 3,  these solutions are
extended to include an axion, or
antisymmetric tensor field. In Section 4, we begin our study
of fermions in this homogeneous time-dependent background. Since the
spatial slices are three spheres ($S^3$),    spinor harmonics on  $S^3$
are reviewed. The coupling to the
gauge field is then included. We find a   fermion zero mode in the background
of the $f=1/2$
saddlepoint solution. It is shown exlicitly that as the gauge field
evolves between vacua, a negative energy level arises from
the Dirac sea, crosses zero energy at the saddlepoint
configuration,  and then enters the positive energy spectrum.

\newsec{Yang-Mills Cosmologies}

In this section,
we consider cosmological solutions with a homogeneous Yang-Mills field as
energy
source {\ref\review{There is a large literature on
cosmological solutions with a Yang-Mills field.
See for example:
J. Tafel,   in {\it Proc. Geometrical and Topological Methods
in Gauge Theories}, ed. J. Harnard and S. Shnider (Springer, Berlin, 1980);
M. Henneaux, {\it Journal Math. Phys.} {\bf 23} (1982) 830;
P. Moniz and J. Mourao, {\cqg} {\bf 8} (1991) 1815;
Y. Verbin and A. Davidson, {\plb} {\bf 229} (1989) 364;
D. Gal'tsov and M. Volkov, {\plb} {\bf 256} (1991) 17;
O. Bertolami, et. al., {\it Int. Journal Mod. Phys.} {\bf 6} (1991) 4149;
M. Bento, O. Bertolami, P. Moniz, J. Mourao, and P. Sa,
{\cqg}{\bf 10} (1993) 285; P. Moniz, J. Mourao, and P. Sa, {\cqg}  {\bf 10}
(1993)
517. }}.  Since   we wish
to match   the interior cosmological solution onto  an exterior solution
initially at rest,  we take  closed spatial
sections because only for $k=1$ is
 there  a point at which
the solution is
momentarily static.    Assume  a Robertson-Walker ansatz
\eqn\rw{\eqalign{
  ds^2 & =  -dt^2 + a^2(t) d\O^2_3\cr
& = a^2 (\eta ) ( -d\e^2 + d\O^2_3)\cr}
}
 with
 $S^3$ spatial sections
  where $d\O^2_3$ is the round metric on $S^3$ and where $\e$
   conformal time with
 ${d\e \over dt} = a^{-1}$. The metric {\rw}
is conformal to the Einstein static universe.
In addition to the spacetime metric, consider an   $SU(2)$ Yang-Mills field
$A_{\m}^a$ where
$a$ is a Lie algebra index with  associated field strength   ${ \bf F} \equiv
d{\bf A} -ie/2 [{\bf A},{\bf A}]$  where ${\bf A} \equiv A_{\m} dx^{\m}$
is the Lie-algebra valued one-form and $e$   the gauge coupling constant.
Under a gauge transformation, $U(x)$, the fields transform as
$A \ra U^{-1}AU  + i/e \, U^{-1}dU $ and $F\ra U^{-1}FU $.
 As we now discuss, since the group manifold
of  $SU(2)$ is $S^3$ and coincides with the spatial sections, there
is a natural ansatz for the gauge field which shares the symmetries of
$d\O^2_3$,
the spatial metric.

Before discussing the gauge field ansatz,
we   review invariant forms on $S^3$.
The natural imbedding of  $S^3$ into  the group manifold of  $SU(2)$ is given
by
\eqn\gmat{
g = \pmatrix{ x^4 + i x^3 & x^2 + i x^1\cr
              - x^2 + i x^1 & x^4 - i x^3\cr}
= x^4 + i x^i \t^i, \quad {\rm det}\, g =  (x^1)^2 +(x^2)^2 +(x^2)^2 +(x^4)^2
=1}
where $\t^i$ are the Pauli spin matrices.
$SO(4)$ transformations are  induced
by its two fold cover $SU(2)_L \times SU(2)_R$
corresponding to multiplication of $g$  on the left and right by $SU(2)$.
Left invariant
one-forms are given by (or more precisely, given  by the pull back
to $S^3$ of)
\eqn\leftform{
{\bf e}^L_i = -i{\rm Tr}\, \t_i g^{-1} dg = 2 (x^4 dx^i - x^i dx^4 + \ep_{ijk}
x^j
 dx^k), \quad i = 1,2,3
}
and obey $d {\bf e}^L_i =  {1\over 2}  \ep_{ijk} {\bf e}^L_j \wedge {\bf e}^L_k
 $.
  Under right multiplication
 the forms ${\bf e}^L_i $ {\leftform} transform
in the adjoint. The right invariant one-forms are ${\bf e}^R_i = i{\rm Tr}\,
\t^i  dg g^{-1} $
and obey  $d {\bf e}^R_i =  {1\over 2}  \ep_{ijk} {\bf e}^R_j
\wedge {\bf e}^R_k  $.
 ${\bf e}^L_i$ and ${\bf e}^R_i$ can be written as
${\bf e}^L_i =  \ep_{ijk} ( {\bf M}_{jk} + *{\bf M}_{jk})$ and
 ${\bf e}^R_i= \ep_{ijk}( {\bf M}_{jk} -* {\bf M}_{jk} )$, the self-dual and
anti-self-dual parts of $ {\bf M}_{ij} =   x^i dx^j -  x^j dx^i$ where
$ *{\bf M}_{ij} \equiv {1\over 2} \ep_{ijkl} {\bf M}_{kl}$.
The left and right invariant vector fields are dual to {\leftform} and given
by
\eqn\vector{\eqalign{
E^L_i &= 1/2( x^4 {\p\,\over \p x^i}- x^i {\p\,\over \p x^4}
 +  \ep_{ijk}  x^j {\p\,\over \p x^k} )\cr
E^R_i &= - 1/2( x^4 {\p\,\over \p x^i}- x^i {\p\,\over \p x^4}
 -  \ep_{ijk}  x^j {\p\,\over \p x^k} ).\cr
}}
$E^L_3$ and $E^R_3$ generate the transformation $(z^1, z^2) \ra (  e^{i\th /2}
z^1,
  e^{i\th /2} z^2)$ and  $(z^1, z^2) \ra (  e^{-i\th /2} z^1,
 e^{i\th /2} z^2)$ respectively  where $z^1 \equiv x^4 + ix^3$ and $z^2 \equiv
x^1 + i x^2$. $L_i = -iE^L_i$ and $R_i = -iE^R_i$   obey the $ SU(2)$
commutation relations $[L_i ,L_j] = i\ep_{ijk} L_k ,$
  $[R_i , R_j] = i\ep_{ijk} R_k ,$ and $[L_i, R_j] =0$.
 The round metric is the pull back to $S^3$ of the flat four dimensional
metric
\eqn\flatfour{
ds^2 =  (dx^1)^2 + (dx^2)^2 +
 (dx^3)^2 + (dx^4)^2  }
and in terms of the matrix $g$ {\gmat}
takes the form
\eqn\metricg{
ds^2 = - {1\over 2} {\rm Tr}\, (g^{-1} dg)^2 = {1\over 4} {\bf e}^L_i
\otimes {\bf e}^L_i
={1\over 4} {\bf e}^R_i \otimes  {\bf e}^R_i .}
 {\metricg} is the bi-invariant metric
with  isometry group $SU(2)_L \times SU(2)_R$
 corresponding  to left
and right translations.

There is a natural ansatz for the gauge field on $S^3$
which  takes the form
\eqn\ym{
A = {i\over e} f (\e ) g^{-1}
{{\d }g }
 = - {1\over e}  f(\e){\bf e}^{L}_i {\t^i\over 2}
.}
 Since   {\ym}  is invariant under left translation and
changes by a global gauge transformation under right translation, the
gauge field is   homogeneous sharing
 the  symmetry group of  the $S^3$ spatial slices.
 $f$ and $\e$ are
dimensionless  while $e^2$ has units of $[M]^{-1} [L]^{-1}$.  The field
strength associated with  {\ym}
is
\eqn\F{
F =  F^i {\t^i\over 2},\quad F^i = -{\dot f\over e} {\d}{\bf \e} \w {\bf
e}^{Li} +  {1\over 2e}f(f-1)    \ep^{i}_{\,\,jk}
 \be^{Lj} \wedge \be^{Lk} .
}
{}From this expression or directly from {\ym}, one observes that $f=0$ and $ 1$
correspond
to  vacuum, or pure gauge field, configurations.   For a general (non-vacuum)
configuration, the Chern-Simons number is given
by the integral  $N_{CS}
= \int_{S^3} \omega_3 $ of the three-form
\eqn\chernsimons{
 {\bf \o}_3 = {e^2 \over 8\pi^2} \tr ( A \wedge dA - {2ie\over 3} A\wedge A
\wedge A)
}
   satisfying  $d {\bf \o}_3 =  {e^2 \over 8\pi^2} \tr F\wedge F$.
  Substituting {\ym} in {\chernsimons}
and using the fact that {\metricg} implies that
${1\over 8}\int {\bf e}_1 \w {\bf e}_2  \w{\bf e}_3  = 2 \pi^2$, the area
of the unit $3$-sphere, one eventually finds
  $N_{CS} = 3f^2 (1-2/3f)$.    The $f=1$ configuration has unit  topological,
or Chern-Simons, number corresponding to the fact that the map $g$ has unit
winding
number.

The gauge field   should satisfy the Yang-Mills field equation
$ * D * F = D_{\m} F^{\m\n} = 0$
where $D \equiv d - ie [A,]$ is the gauge  covariant derivative
and $*$ is the Hodge dual.  Substituting in the
ansatz {\ym}, one obtains an equation of motion for $f$
\eqn\ymeqn{
 {\p^2 f\over \p \e^2}  = - {\p\over \p f} V,\quad V\equiv 2(f^2 -f)^2
.}
The field equation is independent of the scale factor $a (\e)$ since the
Yang-Mills field equation
is conformally invariant. {\ymeqn} describes a particle moving in a double well
potential $V$
with minima at the    pure gauge configurations $f=0$ and $1$. These are the
only pure gauge configurations within this ansatz. There is a conserved
 (dimensionless)
energy type quantity $E = {1\over 2}\dot f^2 + V(f)$
for the equation of motion {\ymeqn}. From the form of the field
strength above, one can identify the electric and magnetic
field strength with the kinetic and potential energy respectively.
 In addition to  the pure gauge configurations,  the $f= 1/2$ configuration at
the top
of the energy barrier  with $E= E_0 \equiv 1/8$ is also a
static solution  which is unstable and has non-zero (purely magnetic)
field strength \ref\hos{Y. Hosotani, {\plb} {\bf 147} (1984) 44.}. This
solution in fact corresponds to
the reduction to three dimensions of
  the  four
dimensional Euclidean Yang-Mills    meron solution {\aff}.
Using the conserved energy, $E$,  one  can integrate to obtain $f(\e)$ in terms
of elliptic functions. For $E=E_0$, there are two solutions starting
and finishing at the top of the energy barrier that  take the simple form
\eqn\ezero{
f(\e) = {1\over 2} \pm { \sqrt{2} \exp{\sqrt{2} \e} \over 1 +
 \exp{2\sqrt{2} \e} } .
}

The  coupling of the Yang-Mills field to gravity is governed by
the   energy-momentum tensor
 \eqn\emym{
T_{\m\n}= \sum_{i=1}^3 (
  F^i_{\m\a} F^{i\,\a}_{\n} - {1\over 4} g_{\m\n} F^i_{\a\b} F^{i\a\b})
{}.
}
 Thus, each term in {\emym} behaves locally like the stress
tensor of a Faraday flux tube with longitudinal tension
and perpendicular pressure. For our ansatz, the three flux
tubes are mutually orthogonal and have equal strengths.
Thus,  the energy-momentum tensor
is of the form of pure radiation $ P = \r /3$ where
the energy density is given by $\r = \r_0/a^4, \;  \r_0  = 12{E\over  e^2}$.
 The solution to Einstein's equation for pure radiation is
  a Tolman universe with scale factor in conformal time
given by
$a (\e) = ({32\pi E  })^{1/2}  l_0\,\sne .$
 Here, $l_0 = {G^{1/2}} {e}^{-1}$ defines  the only length
 scale in
the system. By contrast, $m_0 = G^{-1/2} e^{-1}$  sets
the scale of the mass of the Bartnik-McKinnon solutions.

The scale factor $a (\e )$ has   the familiar $a \sim t^{1/2}$ form for small
time
for a radiation dominated universe.
These solutions describe an expanding universe of radiation in whcih the
Yang-Mills field constituting the radiation oscillates in the potential
well given in {\ymeqn}.
We note that even  though  $f=1/2$ is a static solution to the Yang-Mills
equation
with the field strength, $F^i_{\m\n}$, independent of time, the scale factor is
time dependent.
The lifetime  of the universe
is $\Delta\e = \pi$, or  $\Delta t \sim \sqrt{E} l_0$.
The number of oscillations in the potential
well  in the lifetime
of the universe can be obtained by integrating $f$.
For large $E$,
one has
\eqn\largeE{
\Delta\e = \oint  { df\over \sqrt{2 (E - V(f))} }\propto E^{-1/4},
}
the integral being taken between the turning points.
Thus, the number of oscillations in the lifetime
of the universe grows as $N\sim E^{1/4}$ for large $E$.
As  $E$ tends to $ E_0$ from above, the period of an oscillation
diverges as $\Delta\e  \sim (E-E_0)^{-1/2} .$

 As stated earlier, one motivation for studying these solutions
is that they might serve as a model for the interior of the collapsing
Bartnik-McKinnon sphaleron just as a simple model of the gravitational
collapse of a star is provided by a portion of a $k=1$ pressure free Friedmann
 universe. There are, however, certain difficulties with
modelling the interior by a homogeneous Yang-Mills field.  First, unlike the
case of a pressure free perfect fluid,  even if the stress tensor were
initially homogeneous and isotropic, there is no reason to suppose that it will
remain so. Second, the Bartnik-McKinnon solution
more closely approximates a  shell of Yang-Mills gauge field
rather than a ball, and  numerical studies
{\ref\zhou{Z. Zhou, {\it Helvetica Physica Acta} {\bf 65} (1992) 767.}} suggest
that the shell becomes thinner
as the configuration collapses. Third, since the Yang-Mills field
possesses non-zero pressure, there will be a pressure jump, and therefore,
unlike
pressureless dust  one cannot  simply
  attach the homogeneous interior directly onto the
exterior vacuum.
This last  problem might  be resolved provided  one could smooth out the
boundary between the homogeneous solution and the exterior vacuum.

\newsec{Axion  Field}

In this section, we consider the addition of an axion field, $H_{\m\n\r}$. The
natural ansatz for this field
which shares the symmetries of $S^3$ is given by
\eqn\kr{
{\bf H} = h(\e) {\bf\ep^3 }
 }
where $\ep^3$ is the volume form on $S^3$ with metric $d\O^2_3$.
The equations of motion for $H$ are $d *   H= 0 $ and $ d  H =0$.
{\kr } automatically satisfies the first, while the second implies
that $h$ is a constant. The energy-momentum tensor for $H$ is
\eqn\emkr{
T_{\m\n}=  H_{\m\l\r}H_{\n}^{\,\,\l\r} - {1\over 6} g_{\m\n}
H_{\a\b\g}H^{\a\b\g} .}
Substituting in  {\kr}, one finds that $T_{\m\n}$ is of the form of
a perfect fluid
with $P=\r = h^2/a^6$.

Let us now find the solution for the scale factor, $a$, with
the combined Yang-Mills and axion fields as energy source.
The Friedmann equation is
\eqn\fried{
 {\dot a \over a^2} + {1\over a^2} = {8\pi G \over 3}\r
}
or in conformal time,
\eqn\friedconf{
({da\over d\e})^2 + a^2 = {8\pi G\over 3}  \r a^4
}
where the energy density, $\r$, is the sum
of the Yang-Mills and axion contributions
 $\r = \r_{YM} + \r_{AXION} =12  E/e^2 a^{-4} + h^2 a^{-6}$.
 {\friedconf} can be integrated exactly
to yield the scale factor in
closed form
\eqn\asmscale{\eqalign{
a(\e) &= \bigl ( (\a^2  + \b)^{1/2} \sin 2(\e -\e_0 ) + \a \bigr )^{1/2}
,\quad\cr
 \sin 2\e_0  &= {\a\over \sqrt{\a^2 + \b}}, \quad \a \equiv {16 \pi G E\over
e^2} \quad
   \b\equiv {8\pi G\over 3}  h^2, \quad .\cr}
} The solution still has an expansion followed by a contraction.
Near the singularity, however,  the contribution of $H$ to the energy density
dominates with the scale factor vanishing as $a\sim t^{1/3}$ rather than
$a\sim t^{1/2}$ as in the Tolman universe.

\newsec{Fermions}

We now consider fermions propagating in the Einstein-Yang-Mills
backgrounds discussed above {\ref\rey{Fermions in Euclidean Yang-Mills
wormholes were discussed in S-J. Rey, {\npb}
{\bf 336} (1990) 146; O. Bertolami, ``Wormhole solutions
of    Euclidean Yang-Mills with Fermions,
  13th International Conference on Group
Theoretical Methods,   Moscow (1990).}}.
The  Dirac equation  for a massless Dirac fermion
coupled to gravity and a Yang-Mills field   is given by
\eqn\diraceq{
i\g^{\m} (\nabla_{\m} -i e A^a_{\m} T^a) \P   =0
}
where  $ \nabla_{\m}$ is the covariant derivative
and $T^a$ are   generators for the gauge group.  The massless Dirac equation
is conformally invariant. That is, given a solution  $(\Psi, g_{\m\n}, A_{\m})$
in $d$ spacetime dimensions,
there is  another solution  $(\O^{d-1\over 2} \Psi, \O^{-2} g_{\m\n}, A_{\m})$.
Thus, if $\Psi$ is a solution in the spacetime {\rw}, $\tilde \P = a^{3/2} \P$
is a solution to the Dirac equation on  $S^3 \times R$, the Einstein static
universe (ESU).
The Dirac equation {\diraceq} on ESU takes the form
\eqn\diracham{ (i \g^0 {\partial\,\over \partial \e}
 +i \D3 + e \Aslash T^a )\tilde \Psi =0
}
where $\D3$ is the Dirac operator on $S^3$.
 Consider  the chiral representation for the gamma matrices
\eqn\gammas{
\g^0 = {\pmatrix{0&1\cr 1 & 0\cr}}\quad\quad
\g^i =  {\pmatrix{ 0 & \s^i \cr -\s^i & 0\cr}}\quad\quad
\g^5 = {\pmatrix{ -1 &0\cr \,0& 1\cr}}
}
where $\s^i$ are Pauli spin matrices.

In the absence of the gauge field,
solving {\diracham} reduces to finding  spinor harmonics on $S^3$.
 Before considering spinors, we review  scalar harmonics on $S^3$.
The scalar spherical harmonics  satisfying
 $\nabla^2_3 f_l = \l_l f_l$ are given by
\eqn\sharm{\eqalign{
f_l = \a_{i_1\cdots i_l} n^{i_1} \cdots n^{i_l},&\quad n^i = x^i/r ,
\quad \a^i_{\,\,i i_3 \cdots} =0
\;\;\;({\rm traceless})\cr
\l_l &= - l(l+2) \cr}
}
   where $\nabla^2_3$ is the  Laplacian on $S^3$.
This is easily checked using the form of  the four-dimensional flat space
Laplacian in spherical coordinates
\eqn\lapsph{
\nabla^2_4 = {\pa^2\,\over\pa r^2} + {3\over r}  {\pa \,\over \pa r}
+ {1\over r^2} \nabla^2_3
}
and using  $  \nabla^2_4 (r^l f_l) =0$ from {\sharm}. [In $n$-dimensions, the
harmonics are   also given by {\sharm}, but with eigenvalues $\l_l = -
l(l+n-1)$.]
Since    $SU(2)_L \times SU(2)_R$ is the two-fold cover
of $SO(4)$, $\a_{i_1\cdots i_l}$ can be written in terms of spinor components
 as $\a_{A_1 \dot A_1\cdots A_l \dot A_l}$ where undotted and dotted
components correspond to left and right factors of $SU(2)$ respectively.
The fact that $\a_{i_1\cdots i_l}$ is totally symmetric and traceless
  implies that  $\a_{A_1 \dot A_1\cdots A_l \dot A_l}$ is totally
symmetric in dotted and undotted indices separately. Since the ${k\over 2}$
representation of $SU(2)$
is given by  rank $k$ totally symmetric  $SU(2)$ tensors, the harmonics
transform
  in the $({l\over 2},{l\over 2})$ representation
of $SU(2)_L \times SU(2)_R .$  [Note that
  the harmonics for a squashed sphere form a representation
of  some subgroup of $SU(2)_L \times SU(2)_R$.]
{}From {\vector}, one can show
\eqn\laplacian{
E^L_i E^L_i = E^R_i E^R_i = \nabla^2_3 .}
If one  diagonalizes   $L_3$, and $R_3$ with eigenvalues $m$ and $n$,
then the harmonics can be expressed as
monomials in $z_1, z_2, \bar z_1, \bar z_2$
of degree $l$.  The eigenvalues  $m$ and $n$ of these monomials
are determined from those of   $z_1, z_2, \bar z_1, \bar z_2 .$
They are $m= {1\over 2}, {1\over 2}, -{1\over 2}, -{1\over 2}$ and $n =
-{1\over 2} , {1\over 2}, {1\over 2}, -{1\over 2}$.

Having constructed the scalar harmonics, we now consider
spinor harmonics on $S^3$.
Like the scalar harmonics, spinor harmonics are most simply
expressed in terms of quantities in the four-dimensional
imbedding space. The four-dimensional Euclidean  flat space Dirac operator in
spherical polar coordinates is
\eqn\diracsph{
\paslash = \g^r (     {\pa \,\over \pa r} + {3\over  2} {1\over r})
+ {1\over r} \D3
.}
  The spinor harmonics satisfying
$\D3 \P_l = \l_l \g^r \P_l$
are given by
\eqn\spharm{\eqalign{
\P_l = T^{\a}_{i_1\cdots i_l}& n^{i_1} \cdots n^{i_l},
\quad \g^i_{\a\b} T^{\b}_{i i_2\cdots i_l} =0\cr
\l_l &=  -(l+ {3\over 2}) \cr
}}
where $T^{\a}_{i_1\cdots i_l}$ is constant with $\a$ a Dirac
spinor index
and $i$ a vector index.
 This is  easily checked
 using {\diracsph}  and that {\spharm}
implies $ \paslash (r^l \P_l) =0$.
$T^{\a}_{i_1\cdots i_l}$ can be decomposed into tensors
 $T_{AA_1 \dot A_1\cdots A_l \dot A_l}$ with $l+1$ undotted and $l$ dotted
indices and tensors $T_{\dot AA_1 \dot A_1\cdots A_l \dot A_l}$ with
 $l$ undotted and $l+1$ dotted
indices. {\spharm} implies that they are totally symmetric tensors.
Therefore, the spinor harmonics fall into
 $( {l\over 2} +{1\over 2}, {l\over 2})$
and  $( {l\over 2}  , {l\over 2}+{1\over 2})$ representations of
 $SU(2)_L \times SU(2)_R$ with $l= 0, {1}, \cdots$.
Since one has $\{ \D3, \g^r \} =0 $,  the harmonics with positive eigenvalues
can be obtained by applying
$\g^r$ to {\spharm}.

The spinor harmonics can also be expressed
in terms of an orthonormal frame
$e_i^{\m}$ on $S^3$.  The Dirac operator is then given by
\eqn\sdirac{
i\D3 = i \g^{\m} D_{3\m} , \quad  D_{3\m}= \p_{\m} + {1\over 2}
\o_{\m}^{ij }  \Sigma^{ij},\quad \Sigma^{ij} = - [\g^i, \g^j]/4
}
where $\Sigma^{ij}$ are the Lorentz generators in the spinor
representation with the  Dirac matrices
{\gammas } satisfying $\{ \g^i, \g^j\} =  -2 \de^{ij}$ and  $\o_{\m}^{ij } ,$
the spin connection. As usual
Latin and Greek indices refer to tangent and curved space and
are related by $e_i^{\m}$.
If we let  the left-invariant vector fields {\vector}
 $e_i^{\m} = 2 E_i^{L\m}$
define  the  orthonormal
frame, then, from {\leftform}, the dual
one-forms $e_{\m}^i = {1\over 2}e_{\m}^{Li}$
satisfy  $d {\bf e}_i =   \ep_{ijk} {\bf e}_j\wedge {\bf e}_k  .$ We can now
read off the spin connection $ {\bf \o}_{ij }  =
  \ep_{ijk}  {\bf e}^k$.
 Substituting   the spin connection
and the Dirac matrices in {\sdirac}, one eventually finds
\eqn\dirac{
- i\D3 = {\pmatrix{0& \hat H_0\cr -\hat H_0 &0\cr}}, \quad
\hat H_0 \equiv  4 S \cdot L + 3/2  , \quad S \cdot L \equiv S^i L^i
}
with $S^i = \s^i/2$ and $L^i = -i E_L^i$. $ \hat H_0 $ commutes with $R_i$ and
$J_i \equiv L_i + S_i$. The spinor harmonics
fall into the  $( {l\over 2} +{1\over 2}, {l\over 2}),$ $l=0,1, \cdots$
and  $( {l\over 2} -{1\over 2}, {l\over 2}),$ $l=1,2 \cdots$
representation of  $SU(2)_J \times SU(2)_R$ and can be
 labelled by
$( l,n, j_{\pm}\equiv {l\over 2} \pm {1\over 2}, j_3 ), $  the eigenvalues of
$R^2$, $R_3$,  $J^2$ , and $J_3$. They take the form of products
of  2-component spinors and scalar harmonics.
Since $S\cdot L = (J^2 - L^2 - S^2)/2 = (j(j+1) - {l\over 2} ({l\over 2}
+1) - {3\over 4})/2$, we find
that the spectrum of $\hat H_0$ is
\eqn\spectrum{\eqalign{
\hat H_0  | l,n, j_{\pm}& , j_3 >
= \m_l^{\pm} | l,n, j_{\pm}  , j_3 >, \quad j_{\pm} = {l\over 2}
  \pm{1\over 2} \cr
\m_l^+ & = l+{3\over 2},  \quad l = 0,{1}
\ldots \cr
 \m_l^- & =   -l-{1\over 2},\quad l ={1},2,\ldots
.\cr
 }}
We thus have recovered the spectrum {\spharm}.
The spectrum of $\hat H_0$ is symmetric about zero. The
  lowest lying states in the positive energy spectrum are
$|0,0,{1\over 2}, \pm {1\over 2}> = |\pm>$  $(S_3 |\pm> = {1\over2}|\pm>)$
 with eigenvalue ${3\over 2}$
corresponding to the two components of the constant spinor (with respect to the
left invariant frame). The highest energy states in  the negative
energy spectrum are
   $ |{1\over 2},   {1\over 2}, 0, 0>
=  ( \bar z_1|+> -  z_2 |-> )/ {\sqrt 2}$ and $
|{1\over 2},   -{1\over 2}, 0, 0>
=   ( \bar z_2|+> -  z_1  |-> )/ {\sqrt 2}$
with eigenvalue $-{3\over 2}$.

We now consider the Dirac equation {\diracham}
with a non-zero  gauge field background.
     Let the fermions transform in the fundamental representation of the gauge
group so that  $T^a = \t^a/2 $ with $\t^a$,  the Pauli spin matrices.
{}From {\ym}, we have $e_i^{\m} A_{\m}^a =-2 f/e \de^i_a$ implying  $\s^i
e_i^{\m}A_{\m}^aT^a = -4 f/e S\cdot T$. The full time-dependent Dirac equation
{\diracham} now becomes
\eqn\diractime{
i  {\partial \tilde \psi_{\pm} \over \partial \e} = \mp \hat H \tilde
\psi_{\pm}, \quad \hat H = \hat H_0 + 4f (\e)  S\cdot T
}
with $\hat H_0$  given in {\dirac}   and where we have decomposed
  $\tilde\P = {\pmatrix { \tilde \psi_+\cr \tilde\psi_-\cr}}$
  into  $\tilde \psi_{\pm},$   left and right chiral components
each carrying  two-component Lorentz and internal indices.
For  a  vanishing     gauge field $(f=0)$, $\hat H$  reduces to  $\hat H_0 .$
The   $f=1$ configuration      is pure gauge  with Chern-Simons number unity,
and thus, the spectrum of $\hat H$ should be identical to $\hat H_0$.  Indeed,
 we can redefine $\vec L^{\prime} \equiv\vec L + \vec T$, and the  spectrum  in
the absence of the gauge
 field is recovered.
Since $\hat H_0$ and the interaction $\hat H_I \equiv  4f S\cdot T  $
do not commute,
one cannot in general solve {\diractime}. However, it is possible to do so
within
the  $s$-wave  sector  where $L_i =R_i =0 $. In this sector, there are the
singlet (or hedgehog)
$\chi_1$ and triplet $\chi_2$ states satisfying
$(S+T)^2 \chi_1 =0$ and $(S+T)^2\chi_2 = 2 \chi_2$.
The   time-dependent single and triplet
wave functions are found
from    {\dirac}  and {\diractime}  and by  performing the conformal
transformation back to the Robertson-Walker spacetime {\rey}. One finds
  \eqn\wave{
\eqalign{
\psi^{\pm}_1& = a^{-3/2}\chi_1 \exp \pm{i\int^{\e} E_1 (\e) d{\e}},\quad  E_1
(\e)  = {3\over 2} -3f (\e) \cr
\psi^{\pm}_2& = a^{-3/2} \chi_2 \exp \pm{i\int^{\e}   E_2 (\e) d\e}, \quad  E_2
 (\e ) = {3\over 2}  +f(\e ) \cr
}}

Let us now  consider the adiabatic approximation where
the Hamiltonian, $\hat H ,$ is treated instantaneously.  The energies of the
single and triplet states
are given by
$E_1 (\e )$ and $E_2 (\e)$.
Now as $f$ evolves between the vacua $f=0$ and $f=1$, we see that
 the singlet (hedgehog) state passes from the lowest lying state in the
positive energy spectrum, $E_1 = {3\over 2} ,$
to the highest lying state in the
negative energy spectrum, $E_1 = - {3\over 2} .$
  At the midpoint of
the transition corresponding to the unstable $f=1/2$ solution,
the hedgehog $\chi_1$  is a zero energy mode {\ref\ho{ A. Hosoya and W. Ogura,
{\plb} {\bf 225}
(1989) 117.}}. As pointed out earlier,  $f=1/2$
is the analog of the sphaleron, and like the Bartnik-McKinnon
sphaleron, it has a fermion zero mode.
Since $E_2$ remains positive for $0<f<1$, there is no level crossing
in the triplet state. In fact, since the interaction $2S\cdot T$ is bounded
between $-{3/2} $ and $1/2$,  among all sectors the only level crossing occurs
for the
singlet state. This is consistent with the anomaly equation
according to which if  the Chern-Simons number of the gauge field changes by
unity, there will be
one level crossing in each chiral sector. For Dirac fermions, the level
crossing cancels between chiral sectors.

 We now consider the effect of the antisymmetric tensor field on the fermions.
Consider an  interaction of the form
\eqn\fermionasm{
S = ib \int  \bar\Psi H_{\m\n\r} \g^{\m} \g^{\n} \g^{\r} \Psi  \sqrt {g} d^4x
}
with $b$ having units $[L]^{1/2}[M]^{ -1/2} .$ The Dirac equation now becomes
\eqn\asmdiraceq{
 (i\g^{\m} \nabla_{\m} + e \g^{\m}A^a_{\m} T^a + i b H_{\m\n\r}  \g^{\m}
\g^{\n} \g^{\r} )  \P   =0 .
}
 One can show generally that   a co-closed  axion
field  $({\bf H} = *{\bf d} \phi)$  can be absorbed by a
local chiral transformation on $\Psi$ of the form
\eqn\chiraltrans{
\Psi \ra \exp{(-6ib \g^5 \phi )} \Psi .
}
For the axion field, {\kr}, one finds
that $\phi = \int{h\over a^2}$ and therefore,
from {\wave} and {\chiraltrans}, one finds the time-dependent triplet and
single wave functions
\eqn\hzeromode{\eqalign{
\psi^+_1 &= a^{-3/2} \chi_1  \exp{\pm i\int ( E_1 (\e) -   {6bh \over a^2}})
d\e\cr
\psi^+_2 &= a^{-3/2}\chi_2  \exp{\pm i\int ( E_2 (\e) - {6bh\over a^2}} ) d\e
.\cr
}}

In this paper, we have studied cosmological solutions with a
Yang-Mills field as a potential analytic  model of the gravitational
collapse  of a Bartnik-McKinnon sphaleron. This provides a qualitative picture
of anomalous fermion production and gives some support to the viewpoint
proposed
in the previous paper {\us}. However, it clearly cannot give a detailed
quantitative description. This can, presumably, only be achieved by a numerical
analysis.

\bigbreak\bigskip\bigskip\centerline{\bf Acknowledgements}\nobreak
We would like to thank
Peter Aichelburg, Piotr Bizon, and Paulo Moniz for  discussions.  A.S.
 wishes  to acknowledge   the financial support of the SERC.
\baselineskip=30pt

\listrefs
\end